\documentstyle[epsf,a4,12pt]{article}
\def\be{\begin{equation}}
\def\ee{\end{equation}}

\begin{document}

\begin{titlepage}
\rightline {Si-97-02 \  \  \  \   }

\vspace*{2.truecm}

\centerline{\Large \bf  Monte Carlo Simulation of the Short-time}
\vskip 0.6truecm
\centerline{\Large \bf Behaviour of the Dynamic XY Model }
\vskip 0.6truecm

\vskip 2.0truecm
\centerline{\bf K. Okano$^*$, L. Sch\"ulke, K. Yamagishi$^*$ and 
B.  
Zheng}
\vskip 0.2truecm

\vskip 0.2truecm
\centerline{Universit\"at -- GH Siegen, D -- 57068 Siegen, 
Germany}
\centerline{$^*$ Tokuyama University, Tokuyama-shi, Yamaguchi 754,  
Japan }

\vskip 2.5truecm

\abstract{Dynamic relaxation of the XY model quenched from
a high temperature state to the critical temperature or below
is investigated with Monte Carlo methods. When a non-zero
initial magnetization is given, in the short-time regime
of the dynamic evolution the critical initial increase of
the magnetization is observed. The dynamic exponent $\theta$
is directly determined. The results show that the exponent 
$\theta$
varies with respect to the temperature. Furthermore, it is  
demonstrated
that this initial increase of the magnetization is universal,
i.e. independent of the microscopic details of the initial
configurations and the algorithms.}

\vspace{0.5cm}


\end{titlepage}

\section {Introduction}

Recently much attention has been drawn to the short-time
universal scaling behaviour of critical dynamic systems.
A typical example is the dynamic relaxation of  
magnetic systems quenched from a high
temperature state to the critical temperature.
After a time period enough long in the microscopic sense,
during which the non-universal short wave behaviour is swept away,
the universal scaling behaviour appears.
Such a time period is called the microscopic time
scale $t_{mic}$. $t_{mic}$ is in general
very small compared with the macroscopic time scale,
which is typically $t_{mac} \sim \tau ^{-\nu z}$
or $t_{mac} \sim L ^{z}$. Here $\tau$ is the reduced temperature
and $L$ is the lattice size of the systems, while $\nu$ and $z$ 
are the
critical exponents.
At {\it this macroscopic early stage of the time evolution},
important is that even though
the spatial correlation length of the systems is still
very small in the macroscopic sense, due to
the infinite time correlation length the dynamic systems
already evolve {\it universally}. This is out of the traditional
belief that the universal dynamic scaling emerges only
when the spatial correlation length becomes very big.

One interesting phenomenon is that, if a non-zero magnetization 
is given to the initial state at very high temperature,
at the macroscopic early time
the magnetization surprisingly undergoes 
{\it a critical initial increase} \cite {jan89,hus89}
\begin{equation}
M(t) \sim m_0 \, t^\theta,
\label{e10}
\end{equation}
where $\theta$ is a new critical exponent which is independent
of the static exponents $\beta$, $\nu$ and the dynamic exponent 
$z$.
The exponent $\theta$ is related to the dimension $x_0$ of 
the initial magnetization  $m_0$ by
$\theta=(x_0-\beta/\nu)/z$.
The magnetization continues the increase in a time scale
$t \sim m_0 ^ {-z/x_0}$, then reaches its maximum and
crosses over to the well known long-time universal behaviour.

Numerically the critical increase of the magnetization
has directly been observed \cite  
{li94,sch95,gra95,sch96,cze96,liu95}. 
The critical exponent
$\theta$ was determined to a satisfactory accuracy for the
 two dimensional Ising model and Potts model
\cite {oka97a} as well as the 6-state clock model \cite{cze96}. 
Numerical results show rather clean short-time scaling behaviour.
However, all these models
are relatively simple and the spins locate in discretized
configuration spaces. 
It would be very interesting to investigate
 whether there exists short-time scaling behaviour 
in more complex models.

More important, up to now almost all the numerical simulations
for the short-time dynamics have been done 
with the heat-bath algorithm.
For the heat-bath algorithm 
the microscopic time scale $t_{mic}$ for the two
dimensional Ising and Potts model is not bigger than
one or two Monte Carlo time steps. This is a miracle.
If one Monte Carlo time step is typically the microscopic time 
unit,
 one would expect
that $t_{mic}$ should be around 
$10 \sim 50$ Monte Carlo time steps.
Even though the scaling form in the short-time dynamics
and its applications have extensively been investigated,
it is already overdue to understand universality in the short-time
dynamics, i.e. whether the short-time scaling behaviour
is really independent of the algorithms, lattice types and
other microscopic details.

In a recent paper \cite {oka97a}, the authors have carefully 
analysed
the short-time behaviour of the critical dynamics for the two
dimensional Ising model and Potts model. 
Simulations have been performed with both the heat-bath and the  
Metropolis
algorithm. Indeed, it is only by chance that for the heat-bath
algorithm $t_{mic}$ is negligibly small. Actually
for the Metropolis algorithm, $t_{mic} \sim 30$.
Within the time period up to $t_{mic}$, the dynamic
behaviour for the Metropolis algorithm is very different from
that of the heat-bath algorithm. When $t > t_{mic}$, however,
the dynamic systems with both algorithms present
the same universal behaviour. The measured values of the
 critical exponent $\theta$ are compatible within the 
statistical errors. This is a first step to the
verification
of universality in the short-time dynamics.
More understanding 
is urgent and important.\footnote {Very recently some 
discussions on universality
in short-time dynamics for the two dimensional
Ising model have also been made with respect
to the lattice types and update schemes even though
the critical exponent $\theta$ there was not confidently extracted
due to relative small lattices or some other reasons \cite 
{liu95,rit96a}.}

In this paper, we will report results
for the numerical simulations of the short-time dynamics
of the two dimensional XY model. 
Special attention will be put on universality.
In the next section a short description of the XY model
is given. In the section 3, the numerical data are
presented and conclusions are given in section 4.

\section {The XY model}

The XY model in two dimensions is defined by the Hamiltonian
\begin{equation}
H=K  \sum_{<ij>} \vec S_i \cdot \vec S_j\ ,
\label{e20}
\end{equation}
where $\vec S_i = (S_{i,x},S_{i,y})$ is a planar unit vector at 
site  
$i$ and
the sum is over the nearest neighbours.
 In our notation the inverse temperature
 has been absorbed in the coupling $K$.

The XY model is the simplest statistical system which exhibits
a continuous symmetry. It is known that at a certain critical
temperature the XY model undergoes a Kosterlitz-Thouless phase
transition \cite {kos73,kos74}.
  Near the critical temperature the spatial correlation
length diverges exponentially rather than by a power law
as in the normal second order phase transition.
Below the critical temperature the system remains critical
in the sense that the spatial correlation length is divergent.
No real long range order emerges in the XY model.
The $O(2)$ symmetry of the XY model stays
unbroken in the whole temperature regime.

The XY model is a very important model since it describes the
critical properties of the superfluid helium.
 It is closely related to the
$O(2)$ $\sigma$-model in field theory. Its generalization
such as fully frustrated XY model attracts more and 
more attention \cite {sch89,ram92,lee94}.
The XY model is also a good laboratory to study the 
more general Heisenberg model.

However, due to the exponentially divergent spatial
correlation length, the numerical simulation of the XY model
is very difficult.
Even though some papers can be found concerning the dynamic
properties of the XY model \cite {yur93,rut95}, 
our knowledge about the dynamic
XY model is still very poor. 
In this paper we will
investigate {\it the short-time universal behaviour} of the 
dynamic
XY model. We will concentrate on the critical initial
increase of the magnetization and measure the critical
exponent $\theta$. Special attention will be put on 
universality. Different from the case of the Ising model, 
where the spins take only the values $\pm 1$, 
the spin configuration space for the XY model is a unit circle.
This non-trivial configuration
space allows us to investigate whether the short-time
universal behaviour depends on the microscopic details
of the initial configurations.

\section {Numerical simulations}

Following Janssen, Schaub and Schmittmann's idea \cite {jan89},
we investigate a dynamic relaxation process starting from
an initial state with very high temperature and small
magnetization.  As a first approach to the dynamic
XY model,  we do not consider the effects of the vortices.
The very high initial temperature requests that the spin
at each lattice site is generated independently.
However, the way how to generate a non-zero initial magnetization
in a certain direction is not unique.
A natural way is to introduce an initial
external field, e.g. in the $x$ direction, as it was
used in the numerical simulation of the clock model \cite{cze96}. 
Then the initial
Hamiltonian, i.e. that for generating the initial magnetization,
 can be written as
\begin{equation}
H_{01}=2 h  \sum_{i} S_{i,x}.
\label{e30}
\end{equation}
If we define the magnetization as
\begin{equation}
\vec M(t) = \frac {1}{L^2} \sum_{i} \vec S_{i}
\label{e40}
\end{equation}
with $L$ being the lattice size, the initial
Hamiltonian $H_{01}$ gives an initial magnetization
\begin{equation}
\vec M(0) = (m_0,0) \approx (h,0), \qquad h \to 0.
\label{e50}
\end{equation}
In this paper we are only interested in the case of
small $m_0$.

To prepare the initial state, we update the system described
by  the initial Hamiltonian $H_{01}$
until it reaches equilibrium. Then the generated configurations
of this initial system are used as the initial configurations
of the dynamic system. If the lattice size is infinity,
in each initial configuration an exact value $(m_0,0)$ of the 
initial 
magnetization $\vec M(0)$ is automatically achieved. 
However,
the practical lattice size is finite and the initial magnetization 
$\vec M(0)$
fluctuates around $(m_0,0)$.
This is a kind of extra finite size effect. It
causes a problem in a high precision
measurement. In order to reduce this effect, 
a sharp preparation technique
 has been introduced to adjust
the initial magnetization 
in the numerical simulations of 
the Ising model and the Potts model 
\cite {li94,sch95,oka97a,zhe96a}:
we randomly take one spin on the lattice
and flip the spin if the updated magnetization
comes nearer to the expected value;
we repeat this procedure until the expected initial
magnetization is achieved.
Numerical data show that the sharp preparation technique
improves efficiently the results and especially helps to
obtain better results in relatively small lattices.
 
In the numerical simulation of the XY model,
we also implement the sharp preparation technique.
 However, due to the fact
that spins in the XY model are planar unit vectors,
this procedure becomes
slightly more complicated. We proceed in the following way: 

(i) If the configuration
generated by the initial Hamiltonian $H_{01}$ does give
the value $<S_x>=m$ but not $m_0$, we update a randomly
chosen spin.
If the resulting magnetization $<S_x>$ is nearer to $m_0$,
we accept it otherwise keep the old configuration.
We continue in this way until the difference $|m-m_0| < \delta$
with $\delta$ being a certain given small value.
In our simulations we take $\delta$ to be $2.5$ percent of $m_0$;

(ii) After having adjusted $<S_x>$,
we turn to the magnetization $<S_y>$.
If $|<S_y>| > \delta$, we randomly select a spin
$\vec S_i = (S_{i,x},S_{i,y})$ and change the sign
of $S_{i,y}$. If $|<S_{i,y}>|$
becomes smaller, we accept the new configuration
otherwise keep the old one.
We continue until $|<S_y>| < \delta$.
In this procedure (ii), the value $<S_x>=m$ already prepared in 
(i)
remains unchanged.

After the preparation of the initial configuration,
the system is released to a dynamic evolution
at the critical temperature or below with the Metropolis
or the heat-bath algorithm. We have performed the simulations
for lattice sizes $L=8$, $16$, $32$, $64$ and $128$.
The magnetization is measured up to Monte Carlo time
step $t=150$. The average is taken over $40\ 000$ samples
with  independent
initial configurations
for the lattice size $L\le 64$
and $12\ 000-30\ 000$ samples for the lattice size $L=128$
depending on the initial
magnetization. For smaller $m_0$ we take relatively
large statistics. Errors are estimated by
dividing the data into three or four groups.
In this paper we take the critical temperature
of the XY model from the literature \cite {gup92}, 
$T_c=1/K_c=0.90$.
Unless we explicitly specify, all the discussions below
are assumed to be at the critical temperature.

\begin{figure}[p]\centering
\epsfysize=12cm
\epsfclipoff
\fboxsep=0pt
\setlength{\unitlength}{1cm}
\begin{picture}(13.6,12)(0,0)
\put(1.9,11.0){\makebox(0,0){4}}
\put(1.2,8.0){\makebox(0,0){$\frac {M(t)}{M(0)}$}}
\put(10.8,1.2){\makebox(0,0){$t$}}
\put(9.,8.){\makebox(0,0){\footnotesize$16$}}
\put(12.,8.){\makebox(0,0){\footnotesize$64$,$128$}}
\put(10.,5.5){\makebox(0,0){\footnotesize$L=8$}}
\put(11.5,9.8){\makebox(0,0){\footnotesize$32$}}
\put(0,0){{\epsffile{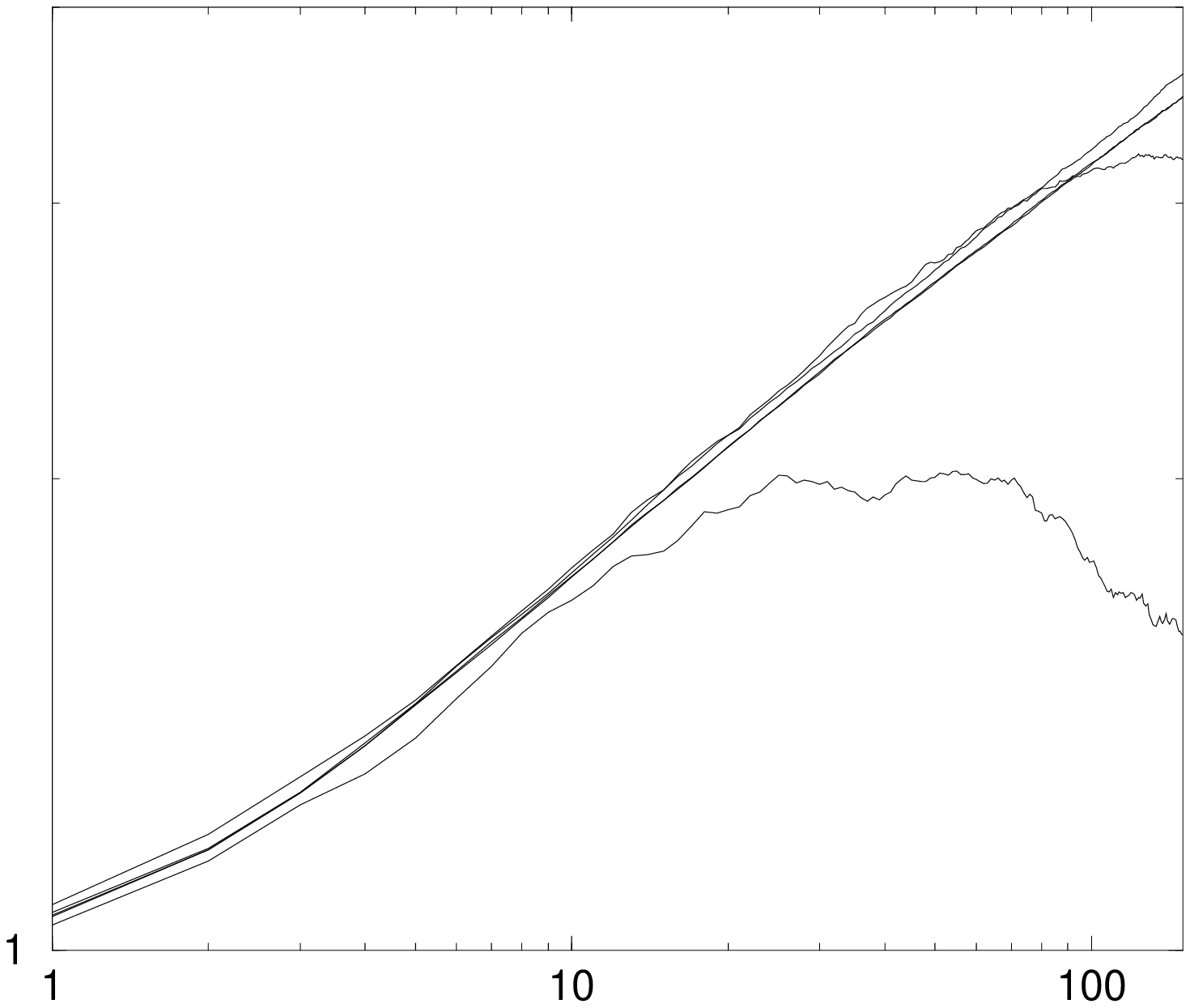}}}
\end{picture}
\caption{ The time evolution of the magnetization for the XY model
with $m_0=0.02$ for different lattice sizes with the Metropolis
algorithm is plotted in double-log scale.
$M(t)$ is the $x$ component of the magnetization
$\vec M(t)$. The sharp preparation technique for the initial
magnetization is adopted.
}
\label{f1}
\end{figure}

In Fig.~\ref{f1}, the time evolution
of the magnetization at the critical temperature
with the initial magnetization
$m_0=0.02$ for the Metropolis algorithm
is displayed for different lattice sizes. In the figure 
$M(t)$ is the $x$ component of the magnetization $\vec M(t)$.
The $y$ component of the magnetization $\vec M(t)$ remains zero
since the initial value is zero.
From the figure we can see that for $L=64$ the finite size
effect is already very small and the curve almost completely
overlaps with that of $L=128$.
We have discussed above that the universal behaviour
appears only after a microscopic time scale $t_{mic}$.
Theoretically one would expect and it is also observed in the 
cases
of the two dimensional Ising model and Potts model that
$t_{mic}$ is in general around $10 \sim 50$.
In Fig.~\ref{f1} one can see this clearly. In the first $30$ time 
steps
there is no universal power law behaviour but after that
it indeed appears. From the slope of the curve one may measure 
the critical exponent $\theta$.

\begin{figure}[p]\centering
\epsfysize=12cm
\epsfclipoff
\fboxsep=0pt
\setlength{\unitlength}{1cm}
\begin{picture}(13.6,12)(0,0)
\put(1.9,11.0){\makebox(0,0){4}}
\put(1.2,8.0){\makebox(0,0){$\frac {M(t)}{M(0)}$}}
\put(10.8,1.2){\makebox(0,0){$t$}}
\put(9.,8.){\makebox(0,0){\footnotesize$16$}}
\put(12.,8.){\makebox(0,0){\footnotesize$32$}}
\put(9.5,5.5){\makebox(0,0){\footnotesize$L=8$}}
\put(11.5,9.8){\makebox(0,0){\footnotesize$64$}}
\put(0,0){{\epsffile{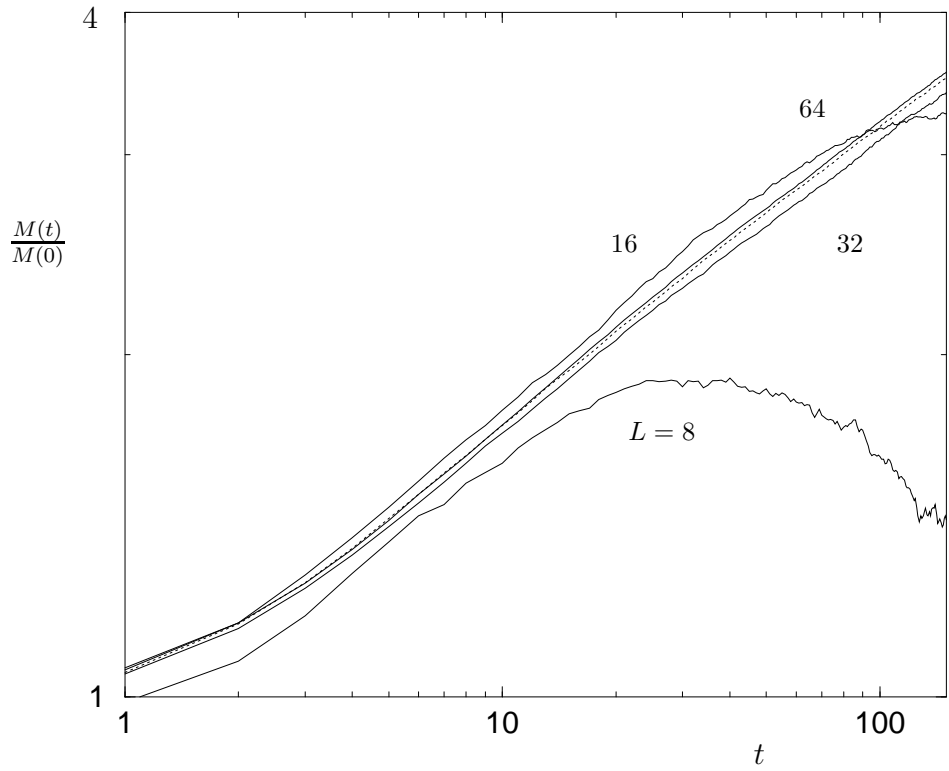}}}
\end{picture}
\caption{ The time evolution of the magnetization for the XY model
with $m_0=0.02$ for different lattice sizes with the Metropolis
algorithm is plotted in double-log scale.
The sharp preparation technique for the initial
magnetization is {\it not} adopted. The dotted line shows the 
magnetization
with the sharp preparation for $L=64$ for comparison.
}
\label{f2}
\end{figure}

How important is the sharp preparation
 of the initial magnetization?
This practically depends on how big the lattice size
and how small the initial magnetization $m_0$ is.
On the other hand, since the exponent $\theta$ is 
defined in the limit $m_0=0$, 
the practically measured 
exponent $\theta$ from the power law behaviour (\ref {e10})
shows in general
some weak dependence on $m_0$ when $m_0$ is finite.
The stronger the dependence of $\theta$ on $m_0$ is,
the more important becomes the sharp preparation.
In Fig.~\ref{f2}, the magnetization {\it without}
a sharply prepared initial magnetization
 is displayed for different lattice sizes.
For comparison, the dotted line shows that with a sharply prepared 
initial magnetization
  for the lattice size $L=64$. 
Comparing Fig.~\ref{f1} and Fig~\ref{f2}
we see that the difference between the curves
with and without the sharp preparation
of the initial magnetization
becomes already quite small 
when the lattice size reaches
$L=64$. Such a small difference 
is also partly due to the quite weak dependence
of $\theta$ on $m_0$, which can be seen later.

With the sharp preparation of the initial magnetization,
the exponent $\theta$ measured from lattice size
$L=64$ and $128$ are $\theta=0.250(1)$ and $0.249(4)$
respectively. Within the statistical errors we already can not
distinguish the results for the lattice size 
$L=64$ and $L=128$. Without the sharp preparation of 
 the initial magnetization,
we get the exponent $\theta=0.252(2)$ for lattice size $L=64$,
which shows a slightly bigger value and fluctuation
compared with that with the sharp preparation of 
the initial magnetization even though
the difference is small. In the following simulations,
the sharp preparation technique is always adopted.

Is the power law scaling behaviour (\ref {e10}) really universal?
Would it depend on the microscopic details of the 
initial configurations, algorithms and lattice types
and so on? In this paper we will show that the power law
behaviour is indeed independent of the microscopic details
of the initial configurations and the algorithms.

In order to generate an initial state
with a non-zero magnetization, using the initial Hamiltonian
$H_{01}$ given in (\ref {e30}) is a natural way but 
by no means unique. An example of alternative methods may be 
the following: in each lattice site, the spin  
orients
towards the pure positive $x$ direction ($S_{i,x}=1,\ S_{i,y}=0$)
with a certain probability, otherwise
randomly. This initial state can be described  
by an initial Hamiltonian
\begin{equation}
H_{02}= \sum_{i} \ln\, ( c_2 \delta (\phi_i) +1).
\label{e60}
\end{equation}
Here the angle $\phi_i$ is defined by
$S_{i,x}=cos\, \phi_i$ and $S_{i,y}=sin\, \phi_i$.
Properly choosing the constant $c_2$ one obtains
the expected initial magnetization $m_0$.
Another possibility is:
assuming that the orientation of initial spins is restricted to 
either
pure $x$ or pure $y$ direction, we give a slightly bigger
probability to generate spins in the positive $x$ direction 
than in others. The corresponding initial
Hamiltonian is 
\begin{equation}
H_{03}= \sum_{i} \ln\, ( c_3 \delta (\phi_i) + \delta (\phi_i-\pi)
   + \delta (\phi_i-\pi/2) + \delta (\phi_i+\pi/2)).
\label{e70}
\end{equation}
It is clear that
the preparation of the initial configurations
given by  $H_{02}$ and $H_{03}$ is
rather simple \cite {li94,sch95,oka97a}.

\begin{table}[h]\centering
$$
\begin{array}{|c|l|l|l|l|}
\hline
        &  \multicolumn{4}{c|} {Metropolis} \\
\hline
  m_0   &   \multicolumn{2}{c|} {0.02}   & \multicolumn{2}{c|} 
{0.01}   
\\
\hline
    & \quad H_{01} &\quad H_{02} &\quad H_{01} &\quad H_{03}\\
\hline
 \theta & .249(4) &  .252(4) & .248(4) & .252(7)\\
\hline
\end{array}
\quad
\begin{array}{|c|l|}
\hline
        &   \multicolumn{1}{c|} {Heatbath}\\
\hline
  m_0   & \quad 0.01  \\
\hline
    & \quad H_{01} \\
\hline
 \theta & .253(5)\\
\hline
\end{array}
$$
\caption{
 The exponent $\theta$ measured for lattice size $L=128$
with different types of initial configurations and algorithms.
}
\label{T1}
\end{table}

\begin{table}[h]\centering
$$
\begin{array}{|c|l|l|l|l|l|}
\hline
 T  &\quad 0.90 &\quad 0.86 &\quad 0.70 &\quad 0.50 &\quad 0.30\\
\hline
\theta & 0.250(1) & 0.264(5) & 0.287(3) & 0.283(4) & 0.282(1)\\
\hline
\end{array}
$$
\caption{ The exponent $\theta$ measured for different 
temperatures
with the Metropolis algorithm. The lattice size is $L=64$.}
\label{T2}
\end{table}

\begin{figure}[p]\centering
\epsfysize=12cm
\epsfclipoff
\fboxsep=0pt
\setlength{\unitlength}{1cm}
\begin{picture}(13.6,12)(0,0)
\put(1.9,11.0){\makebox(0,0){0.1}}
\put(1.2,8.0){\makebox(0,0){$M(t)$}}
\put(10.8,1.2){\makebox(0,0){$t$}}
\put(0,0){{\epsffile{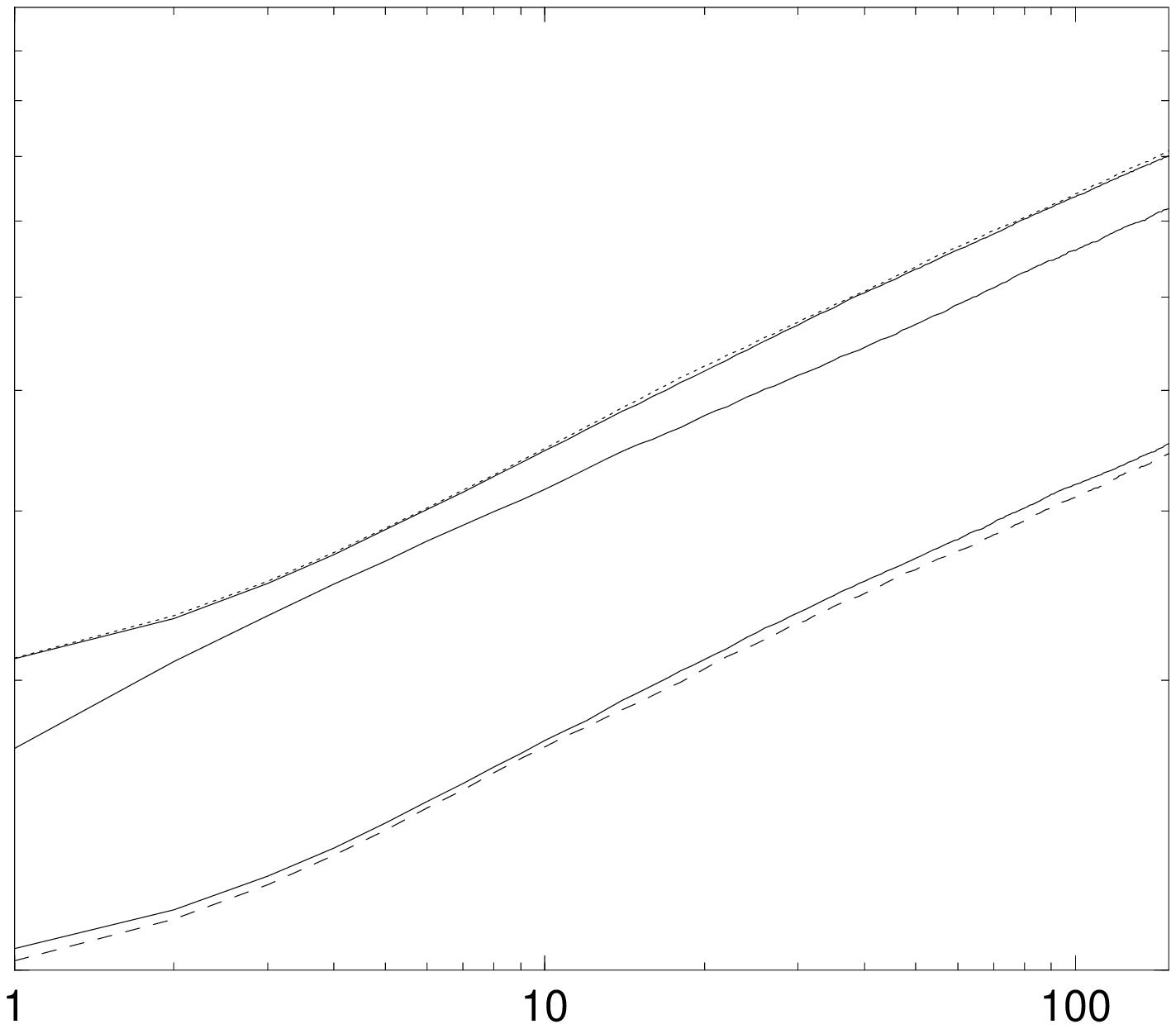}}}
\put(1.8,2.0){\makebox(0,0){0.01}}
\end{picture}
\caption{ The time evolution of the 
 magnetization for the XY model
for lattice size $L=128$ with different types of initial
configurations and 
algorithms is plotted in double-log scale.
The sharp preparation technique for the initial
magnetization is adopted.
 The solid lines are those
obtained with the initial Hamiltonian $H_{01}$. 
The solid lines above and below correspond to $m_0=0.02$ and 
$0.01$
with the Metropolis algorithm. The solid line in between is that
for the heat-bath algorithm with $m_0=0.01$.
The dotted line is for the Metropolis algorithm
with $m_0=0.02$ prepared with $H_{02}$ and the dashed line
is for the Metropolis algorithm
with $m_0=0.01$ prepared with $H_{03}$.}
\label{f3}
\end{figure}

 In Fig.~\ref{f3} the time dependent magnetization
 for different types of initial configurations is plotted
for the lattice size $L=128$.
The solid lines above and below are the results for an
initial magnetization $m_0=0.02$ and $m_0=0.01$
generated from $H_{01}$. The dotted line is the magnetization 
with $m_0=0.02$ from $H_{02}$, and the dashed line
corresponds to that of $m_0=0.01$ with $H_{03}$.
In Table~1 the corresponding $\theta$ measured
in a time interval $[40,150]$ are listed.
We see that all three initial Hamiltonians give
 almost the same results. The difference of the
 initial configurations
 is swept away in more or less one Monte Carlo time step.
Furthermore, the difference
of $\theta$ measured from different initial magnetizations $m_0$
is quite small and already within the statistical errors.
Therefore the extrapolation of $\theta$ to the limit
$m_0=0$ is not necessary here. This is also
one of the reasons why the results with and without
the sharp preparation of
 the initial magnetization are not so different.

Before we continue the discussions of the numerical data,
we would like to make some comments here.
For years it is believed that two exponents $\beta$ and $\nu$ 
sufficiently describe the critical scaling properties of
most magnetic systems in equilibrium and the dynamic scaling
properties can be described by the dynamic
exponent $z$. An essential point in the short-time dynamic scaling
exists in the claim that an {\it independent} critical
exponent $x_0$ (or $\theta$) should be introduced to specify
 the dependence of the scaling behaviour
on the initial magnetization.
In principle, however, there are other 
 choices for the scaling  
variable.
For example, the initial magnetic field $h$ in $H_{01}$
may also be used.
Some discussions concerning what is a better choice of
the scaling variable has recently been made \cite
{zhe96,zhe96a,rit96b}.
In our numerical simulations, we have demonstrated
that a non-zero initial magnetization $m_0$ can be realized
in different ways, by introducing either an initial
magnetic field with $H_{01}$ or some other initial 
systems described by $H_{02}$ or $H_{03}$. However, 
the exponent $\theta$ is the same for
different types of initial configurations. Therefore
these different ways may be considered as the microscopic
details for the initial state. Introducing 
an initial magnetic field is only one possibility
to generate a non-zero initial magnetization.
In this sense,
the scaling variable $m_0$ seems to be more general.

\begin{figure}[p]\centering
\epsfysize=12cm
\epsfclipoff
\fboxsep=0pt
\setlength{\unitlength}{1cm}
\begin{picture}(13.6,12)(0,0)
\put(1.2,8.0){\makebox(0,0){$\theta (t)$}}
\put(10.8,1.2){\makebox(0,0){$t$}}
\put( 6.5, 4.5){\makebox(0,0){\bf\large $\times$\ \ Metropolis}}
\put( 6.5, 3.5){\makebox(0,0){\bf\large $\Box$\ \ heat-bath}}
\put(0,0){{\epsffile{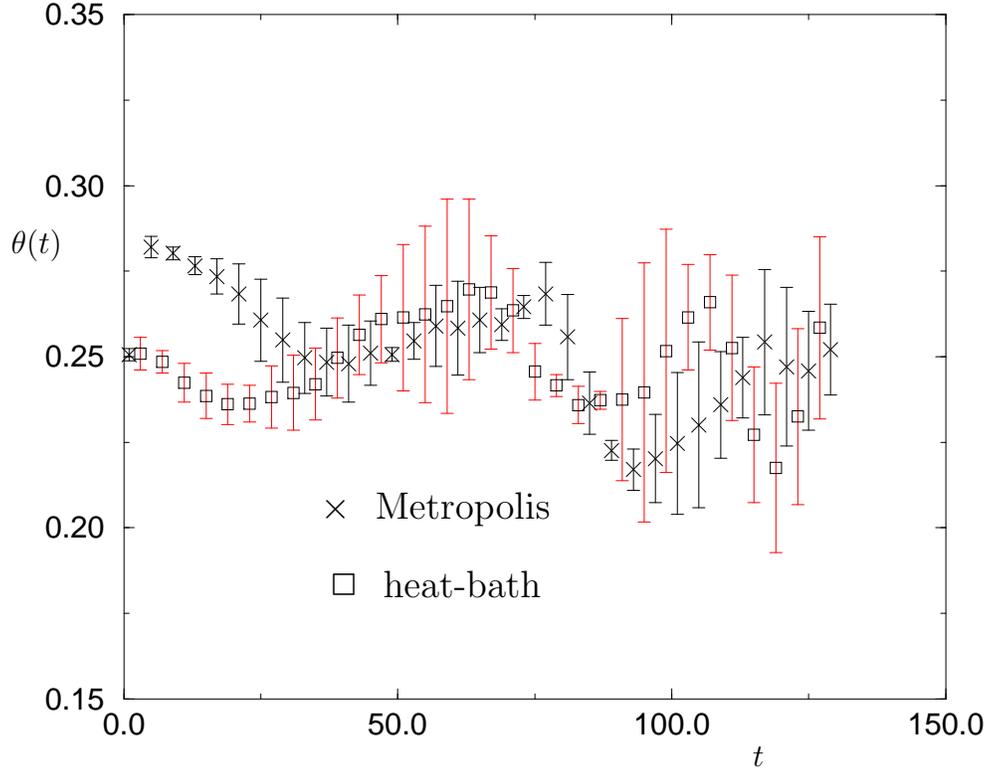}}}
\end{picture}
\caption{The exponent $\theta$ measured as a function of time $t$
for both the heat-bath and the Metropolis algorithm with
$m_0=0.01$ and $L=128$.
$\theta (t)$ is obtained in a time interval $[t,t+20]$.
The sharp preparation technique for the initial
magnetization is adopted.}
\label{f4}
\end{figure}

Now let us come back to our discussions of the numerical results.
In Fig.~\ref{f3} the time evolution of the magnetization
with $m_0=0.01$ with the heat-bath algorithm is also displayed
by the solid line in between.
In the first $20 \sim 30$ time steps, its behaviour is
different from that with the Metropolis algorithm.
After that, however, as in the case of the Metropolis 
algorithm it stabilizes to the universal power law
behaviour. To see this more clearly, in Fig.~\ref{f4} we have
plotted
the exponent $\theta$ as a function of the time $t$
for both the heat-bath and the Metropolis algorithm.
The exponent $\theta$ at time $t$ is measured as the slope
of the curve in a time interval $[t, t+20]$.
Error bars are estimated by dividing the total sample
into three groups. After a microscopic time scale 
$t_{mic} \sim 20 - 30$, the exponents $\theta$ for 
both the heat-bath and the Metropolis algorithm
overlap each other.
The relatively small error bars for the exponent $\theta$
at certain time periods may come from the fact
that the errors are estimated from only three
groups of the data.
The final values for $\theta$ are given in Table~1.
 The results for 
both the heat-bath and the Metropolis algorithm
are also consistent within the statistical errors.
All these results strongly support universality
in the short-time dynamics.

\begin{figure}[p]\centering
\epsfysize=12cm
\epsfclipoff
\fboxsep=0pt
\setlength{\unitlength}{1cm}
\begin{picture}(13.6,12)(0,0)
\put(1.9,11.0){\makebox(0,0){8}}
\put(1.2,8.0){\makebox(0,0){$\frac {M(t)}{M(0)}$}}
\put(10.8,1.2){\makebox(0,0){$t$}}
\put(9.,8.5){\makebox(0,0){\footnotesize$T=0.30$}}
\put(10.,4.5){\makebox(0,0){\footnotesize$T=0.90$}}
\put(0,0){{\epsffile{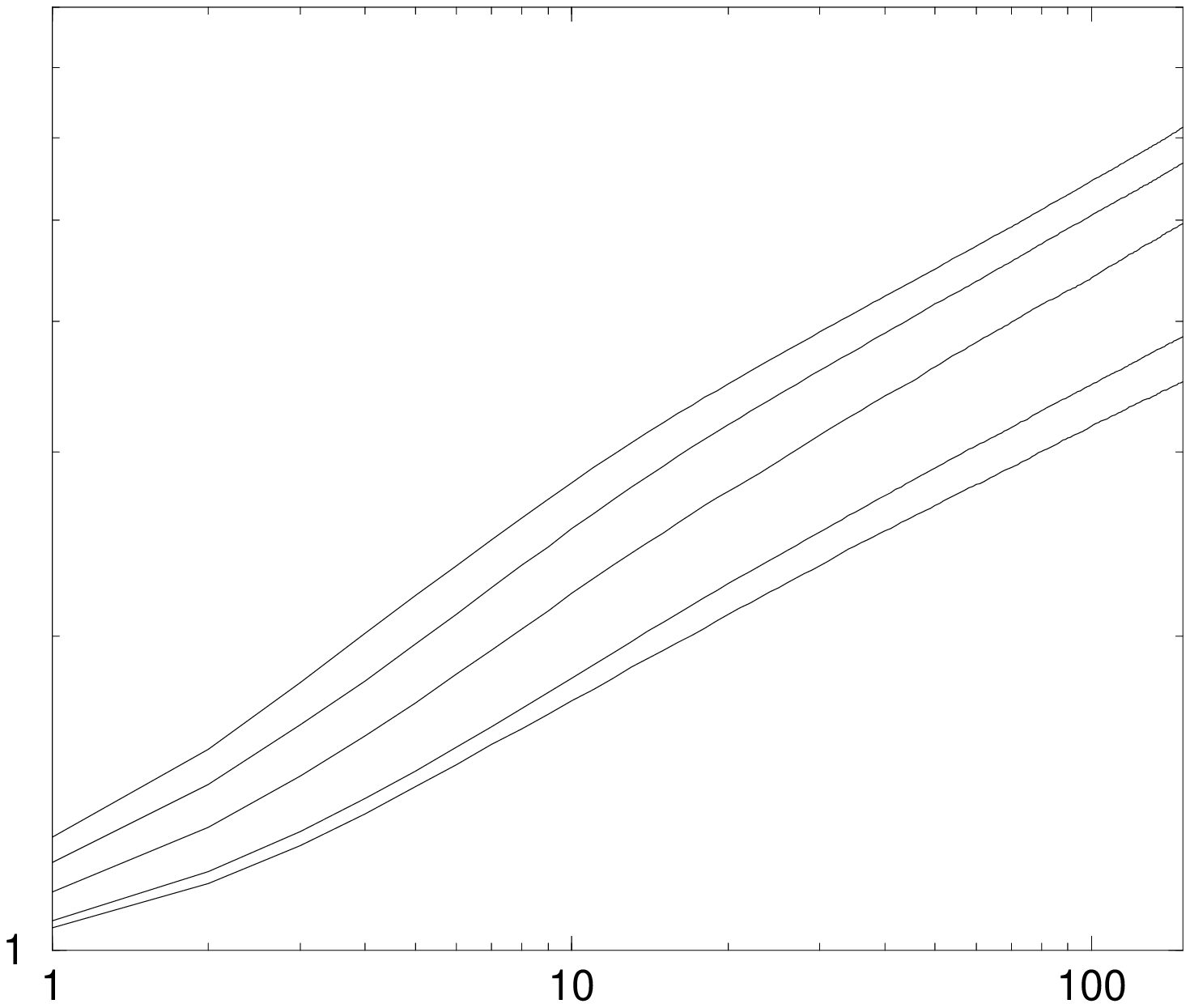}}}
\end{picture}
\caption{ The time evolution of the magnetization
 for lattice size $L=64$ and different temperatures
 with the Metropolis
algorithm is plotted in double-log scale.
The sharp preparation technique for the initial
magnetization is adopted.
 The  
temperature parameters
are $T=0.90$, $0.86$, $0.70$, $0.50$ and $0.30$ (from below).
}
\label{f5}
\end{figure}

Finally we have also performed the simulations
with the Metropolis algorithm
for the temperature below the critical temperature.
Since the XY model remains critical,
a similar scaling form is expected.
In Fig.~\ref{f5}, the magnetization for $L=64$ and
different temperatures
is plotted versus time $t$ in double-log scale.
For the temperatures $T=0.90$ and $0.86$ the initial magnetization  
is $m_0=0.02$.
For the temperatures $T=0.70$, $0.50$ and $0.30$ the initial 
magnetization  
is $m_0=0.01$.
As before, the weak dependence of the exponent $\theta$
on $m_0$ has not been considered since it is within our 
statistical  
errors.
The exponent $\theta$ measured in a time interval $[40,150]$ for  
different
temperatures
is listed in Table~2. It is
known that in the equilibrium
the critical exponents in general depend on
the temperature. Here we see the exponent $\theta$ also
varies with respect to the temperature.
This situation is similar to that of the clock model \cite{cze96}.

\section {Conclusions}

We have numerically investigated
the short-time behaviour of the dynamic relaxation
of the two dimensional XY model at the critical temperature and  
below,
starting from an initial state with a very high temperature
and non-zero magnetization. The critical initial increase of 
the magnetization is observed and the exponent $\theta$
is determined. The results show that
as the temperature decreases, the exponent $\theta$ first
increases rather rapidly and then decreases slowly.
 The independence of the scaling behaviour
on the microscopic details of the initial configurations
and the algorithms is demonstrated and
the microscopic time scale $t_{mic} \sim 30$.
 Universality
in the short-time dynamics is confirmed.
Further extension of this work remains important, such as the
determination of the critical temperature and the static exponents
from the short-time dynamics and an investigation of the effects 
of
the vortices. 

\newpage

%\bibliographystyle{stybase/pr_np}
%\bibliography{stybase/ising}

%\bibliographystyle{tex:inputs/misc/pr_np}
%\bibliography{/ising} 

\end{document}